\documentclass[conferences]{IEEEtran}

\usepackage{mathtools}

\usepackage{floatflt}

\usepackage{color}
\usepackage{cite}

\usepackage{graphicx}
\usepackage{amsmath}
\usepackage{amsthm}
\usepackage{amssymb}

\usepackage{verbatim}

\usepackage{psfrag,array}

\usepackage{subfigure}

\usepackage{stfloats}

\usepackage{todonotes}

\usepackage{amsmath}
\usepackage{epstopdf}
\usepackage{algorithmic}
\usepackage{algorithm}

\newcommand{\p}[1]{\mathop{\mbox{\it p} } }

\renewcommand{\vec}[1]{\ensuremath{\boldsymbol{#1}}}
\newcommand{\be}{\begin{equation}}
\newcommand{\ee}{\end{equation}}
\newcommand{\ba}{\begin{array}}
\newcommand{\ea}{\end{array}}
\newcommand{\bea}{\begin{eqnarray}}
\newcommand{\eea}{\end{eqnarray}}
\newcommand{\bean}{\begin{eqnarray*}}
\newcommand{\eean}{\end{eqnarray*}}
\newcommand{\argmax}{\mathop{\arg\max}}

\newcommand{\rmh}{^{\rm H}}

\definecolor{white}{rgb}{1,1,1}

\usepackage{nopageno}

\begin{document}

\title{User Assignment with Distributed Large Intelligent Surface (LIS) Systems}
\author
{
Sha Hu, Krishna Chitti, Fredrik Rusek, and Ove Edfors\\
Department of Electrical and Information Technology, Lund University, Lund, Sweden\\ \{firstname.lastname\}@eit.lth.se.}

\maketitle

\thispagestyle{empty}

\vspace*{-8mm}

\begin{abstract}
In this paper, we consider a wireless communication system where a large intelligent surface (LIS) is deployed comprising a number of small and distributed LIS-Units. Each LIS-Unit has a separate signal process unit (SPU) and is connected to a central process unit (CPU) that coordinates the behaviors of all the LIS-Units. With such a LIS system, we consider the user assignments both for sum-rate and minimal user-rate maximizations. That is, assuming $M$ LIS-Units deployed in the LIS system, the objective is to select $K$ ($K\!\leq\!M$) best LIS-Units to serve $K$ autonomous users simultaneously. Based on the nice property of effective inter-user interference suppression of the LIS-Units, the optimal user assignments can be effectively found through classical linear assignment problems (LAPs) defined on a bipartite graph. To be specific, the optimal user assignment for sum-rate and user-rate maximizations can be solved by linear sum assignment problem (LSAP) and linear bottleneck assignment problem (LBAP), respectively. The elements of the cost matrix are constructed based on the received signal strength (RSS) measured at each of the $M$ LIS-Units for all the $K$ users. Numerical results show that, the proposed user assignments are close to optimal user assignments both under line-of-sight (LoS) and scattering environments.
\end{abstract}

\section{Introduction}

Large Intelligent Surface (LIS) is a newly proposed wireless communication system \cite{HRE171, HRE172} that can be seen as an extension of massive MIMO\cite{M10, MM12, MM14} systems, but scales up beyond the traditional antenna-array concept. As envisioned in \cite{HRE171, HRE172}, a LIS allows for an unprecedented focusing of energy in three-dimensional space, remote sensing with extreme precision and unprecedented data-transmissions, which fulfills visions for the 5G communication systems \cite{AZ14} and the concept of Internet of Things \cite{IoT} where massive connections and various applications are featured. In \cite{HRE171}, fundamental limits on the number of independent signal dimensions are derived under the assumption of a single deployed LIS with infinite surface-area. The results reveal that with matched-filtering (MF) applied, the inter-user interference of two users at the LIS is close to a sinc-function, and consequently, as long as the distance between two users are larger than half the wavelength, the inter-user interference is negligible. 

\begin{figure}[t]
\begin{center}
\vspace*{-3mm}
\hspace*{0mm}
\scalebox{0.41}{\includegraphics{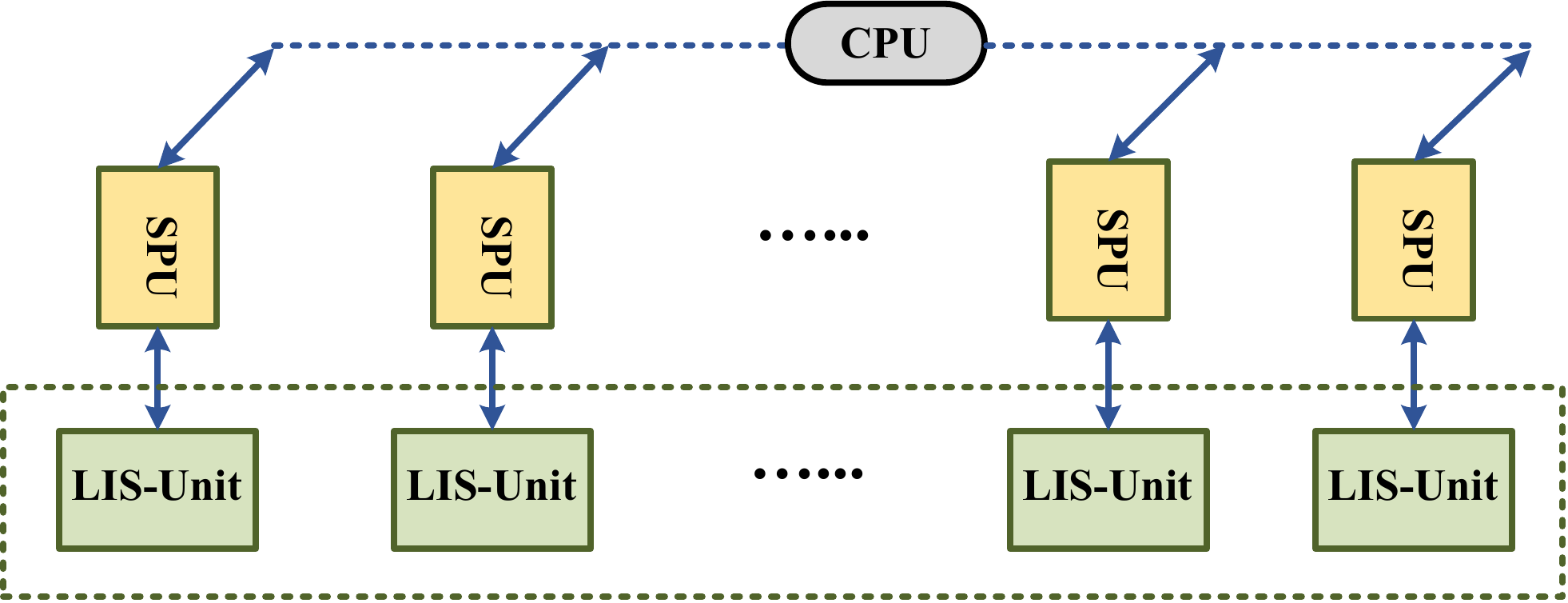}}
\vspace*{-2mm}
\caption{\label{fig1}The diagram of a distributed LIS-communication system. In practical systems, it can also correspond to a centralized deployment of a single large LIS, but with partitioning the LIS into a number of independent segments.}
\vspace*{-7mm}
\end{center}
\end{figure}

In practical deployments, compared to a centralized deployment of a single large LIS, a LIS system that comprises a number of small LIS-Units such as in Fig. \ref{fig1} has several advantages. Firstly, the surface-area of each LIS-Unit can be sufficiently small which facilitates flexible deployments and configurations. For instance, LIS units can be added, removed, or replaced without significantly affecting system design. Secondly, each LIS-Unit can have a separate signal process unit (SPU) which makes cable and hardware synchronizations \cite{JT17} simpler. Thirdly, a distributed LIS-system can provide robust data-transmission and cover a wide area as different LIS-Unit scan be deployed apart from each other. With all these advantages, in this paper we consider optimal user assignments both for sum-rate and minimum user-rate maximizations in a distributed LIS-system that comprises $M$ LIS-Units. The target is to select $K$ ($K\!\leq\!M$) LIS-Units to serve $K$ autonomous users, with each LIS-Unit serving a user separately and simultaneously. 

Firstly, following the work in \cite{HRE171} we show that, with a rather small surface-area, each LIS-Unit is effective in inter-user interference suppression. Hence, the achieved user-rate at each LIS-Unit can be evaluated by the received signal strength (RSS) for each of the $K$ users. Secondly, by specifying a cost matrix whose elements are the RSS at each LIS-Unit for each user, we can construct a bipartite graph between different users and LIS-Units. With such a bipartite graph, the optimal user assignment for sum-rate maximization can be transferred to a linear sum assignment problem (LSAP), while the minimum user-rate maximization can be transferred to a linear bottleneck assignment problem (LBAP), respectively. Then, the transferred linear assignment problems (LAPs) are solved through the well-known Kuhn-Munkres algorithm \cite{K55, KMA}, and Threshold algorithm \cite{THA}, respectively. Both algorithms have time complexities close to $\mathcal{O}(KM^2)$ and are guaranteed to converge to optimal solutions \cite{BL71, G06, DM99}. Lastly, we show through numerical results that, the proposed user assignments solved with the LAPs are close to the optimal schemes both for considered line-of-sight (LOS) and scattering environments.

\section{Received Signal Model with LIS}

\subsection{Narrowband received signal model at each LIS-Unit}

\begin{figure}[t]
\begin{center}
\vspace*{-41mm}
\hspace*{6mm}
\scalebox{0.64}{\includegraphics{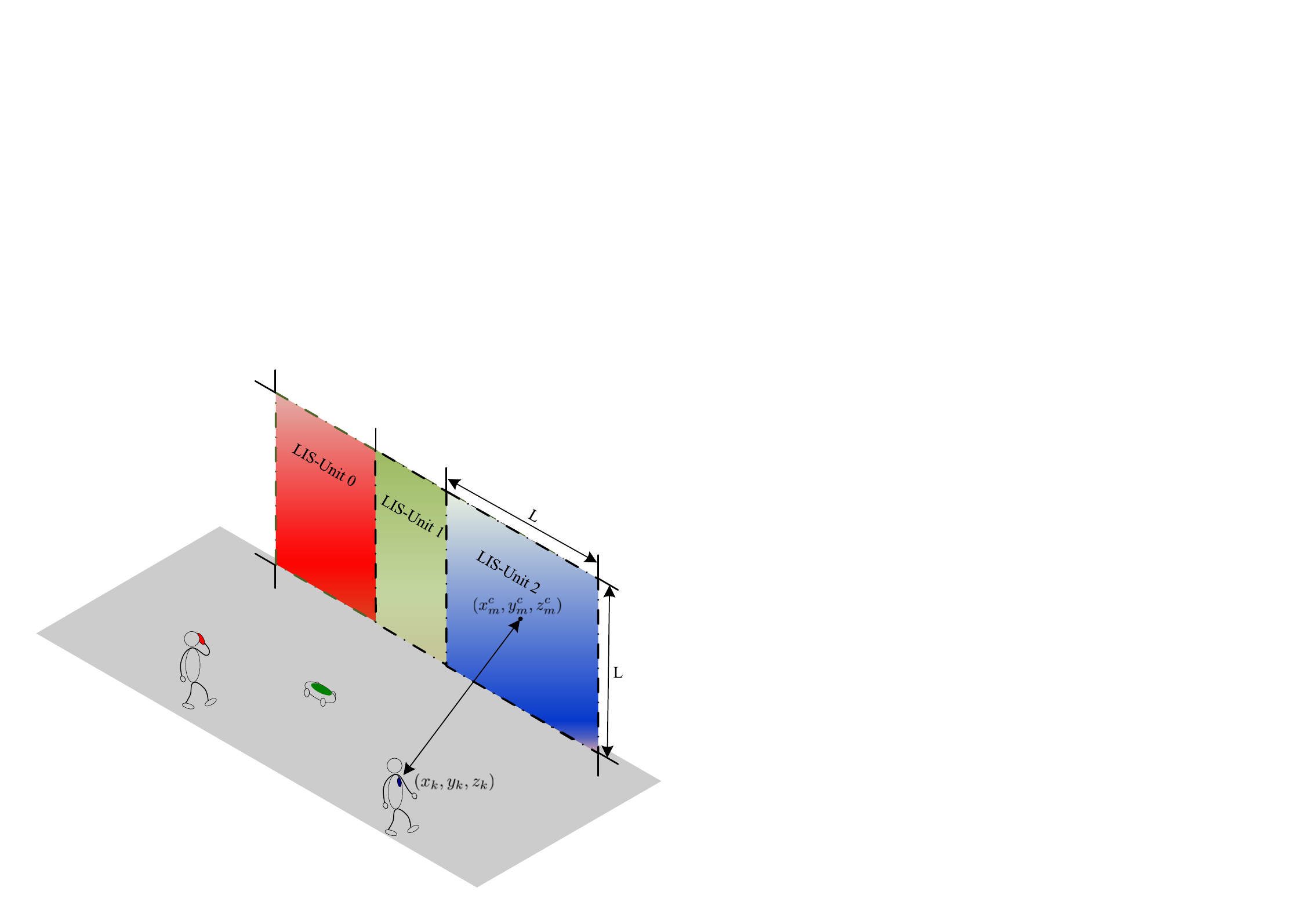}}
\vspace*{-11mm}
\caption{\label{fig2}An LoS scenario where three users communicates to a LIS with separate LIS-Units, with each LIS-Unit serving an individual user simultaneously.}
\vspace*{-7mm}
\end{center}
\end{figure}

We consider the transmission from $K$ autonomous single-antenna users located in a three-dimensional space to a two-dimensional LIS deployed on a plane as depicted in Fig. \ref{fig2}. Expressed in Cartesian coordinates, the center of the $m$th LIS-Unit is located at $(x_m^{\mathrm{c}},y_m^{\mathrm{c}},z_m^{\mathrm{c}})$ with $z_m^{\mathrm{c}}\!=\!0$, while users are located at $z\!>\!0$ and arbitrary $x$, $y$ coordinates. For analytical tractability, we assume a perfect LoS propagation and the case in scattering environments is similar. The $k$th terminal located at $(x_k,y_k,z_k)$ transmits data symbol $a_{k}$ with power $P_k$, which is assumed to be a Gaussian variable with zero-mean and unit-variance, and independent over index $k$. Denote $\lambda$ as the wavelength and consider a narrowband system where the transmit times from users to the LIS are negligible compared to symbol period which yields no temporal interference. Following \cite{HRE171, HRE172}, the effective channel $s_{x_k,\,y_k,\,z_k}(x,y)$ for the $k$th user at position $(x, y, 0)$ at the $m$th LIS-Unit can be modeled as
\bea \label{md1} s_{x_k,\,y_k,\,z_k}^m(x,y)=\frac{\sqrt{ z_k}}{2\sqrt{\pi}\eta^{\frac{3}{4}}}\exp\!\left(\!-\frac{2\pi j\sqrt{\eta_{k,m}}}{\lambda}\right)\!,   \eea
where the metric
\bea \eta_{k,m}=(x_k-x)^2+(y_k-y)^2+z_k^2.\eea 

Based on (\ref{md1}), the received signal at location $(x, y, 0)$ of the $m$th LIS-Unit comprising all $K$ users is
\bea \label{rxyt} r_m(x,y) = \sum_{k=0}^{K-1}\sqrt{P_k}s_{x_k,y_k,z_k}^m(x,y)a_{k} +n_m(x,y),\eea
where $n_m(x,y)$ is AWGN. Given the received signal (\ref{rxyt}) across the LIS-Unit, the discrete received signal after the MF process corresponding to the $k$th user equals
 \bea \label{rkm} r_{m,k} =\sum_{\ell=0}^{K-1}\sqrt{P_kP_\ell} \phi_{k,\ell}^m a_{\ell}+w_{m,k}, 
  \eea
where $w_k[m]$ is the effective colored noise after MF which has a zero-mean and satisfies $\mathbb{E}(w_{m,k}w_{m,k}\rmh)\!=\!N_0\phi_{k,k}$, and the coefficient $\phi_{k,\ell}$ is computed as
 \bea \label{phi} \phi_{k,\ell}^m =\iint\limits_{(x,\,y)\in\mathcal{S}_m}s_{x_\ell,y_\ell,z_\ell}^m(x,y) s^{m,\ast}_{x_k,y_k,z_k}(x,y)\mathrm{d}x \mathrm{d}y. \eea
The variable $\phi_{k,\ell}$ denotes the RSS for the $k$th user with $\ell\!=\!k$, and the inter-user interference between the $k$th and $\ell$th users with $\ell\!\neq\!k$, respectively.

\subsection{Interference Suppression with the LIS-Unit}
Next we evaluate the interference suppression property at each of the LIS-Units. As shown in \cite{HRE171}, when the surface-area of the LIS is infinitely large, two users can be almost perfectly separated without interfering each other after the MF process. However, in practical deployments the surface-area of each LIS-Unit is limited. Therefore, it is of interest to investigate the interference suppression ability for a LIS-Unit with a finite surface-area.

Without loss of generality, we consider two users located in front of a square-shaped LIS-Unit whose center is located at position $x\!=\!y\!=\!z\!=\!0$. Then, the inter-user interference ($\ell\!\neq\!k$) according to (\ref{phi}) equals
 {\setlength\arraycolsep{0pt}\bea \label{phi1} \phi_{k,\ell }^m&=&\int\limits_{-L/2}^{L/2}\!\int\limits_{-L/2}^{L/2}\!\!s_{x_\ell,y_\ell,z_\ell}(x,y) s^\ast_{x_k,y_k,z_k}(x,y)\mathrm{d}x \mathrm{d}y \notag \\
 &=&\int\limits_{-L/2}^{L/2}\!\int\limits_{-L/2}^{L/2} \!\!\!\frac{\sqrt{ z_k z_\ell}}{4\pi(\eta_k\eta_\ell)^{\frac{3}{4}}}\exp\!\left(\!\frac{2\pi j(\sqrt{\eta_{k,m}}\!-\!\sqrt{\eta_{\ell,m}})}{\lambda}\right)\!\mathrm{d}x \mathrm{d}y. \notag \\ \eea}
\hspace{1mm}In \cite{HRE171} we show that, under the condition, $z_k\!=\!z_\ell$, $L\!=\!\infty$ and $\lambda$ is sufficient small, $\phi_{k,\ell }$ only depends on the distance $d$ between two user that equals
\bea d=\sqrt{(x_k-x_\ell)^2+(y_k-y_\ell)^2+(z_k-z_\ell)^2},  \eea
and can be well approximated by a sinc-function. Such a fact leads to an important observation that, as long as two users are located at least $\lambda/2$ away from each other, $\phi_{k,\ell }$ is negligible. However, for a finite $L$, closed-form expression of (\ref{phi1}) seems out of reach and we calculate (\ref{phi1}) through numerical computations. As shown next, we can also see that with a rather small $L$, the LIS-Unit is still quite efficient in suppressing the inter-user interference.

\begin{figure}[t]
\begin{center}
\vspace*{-0.5mm}
\hspace*{-2mm}
\scalebox{0.285}{\includegraphics{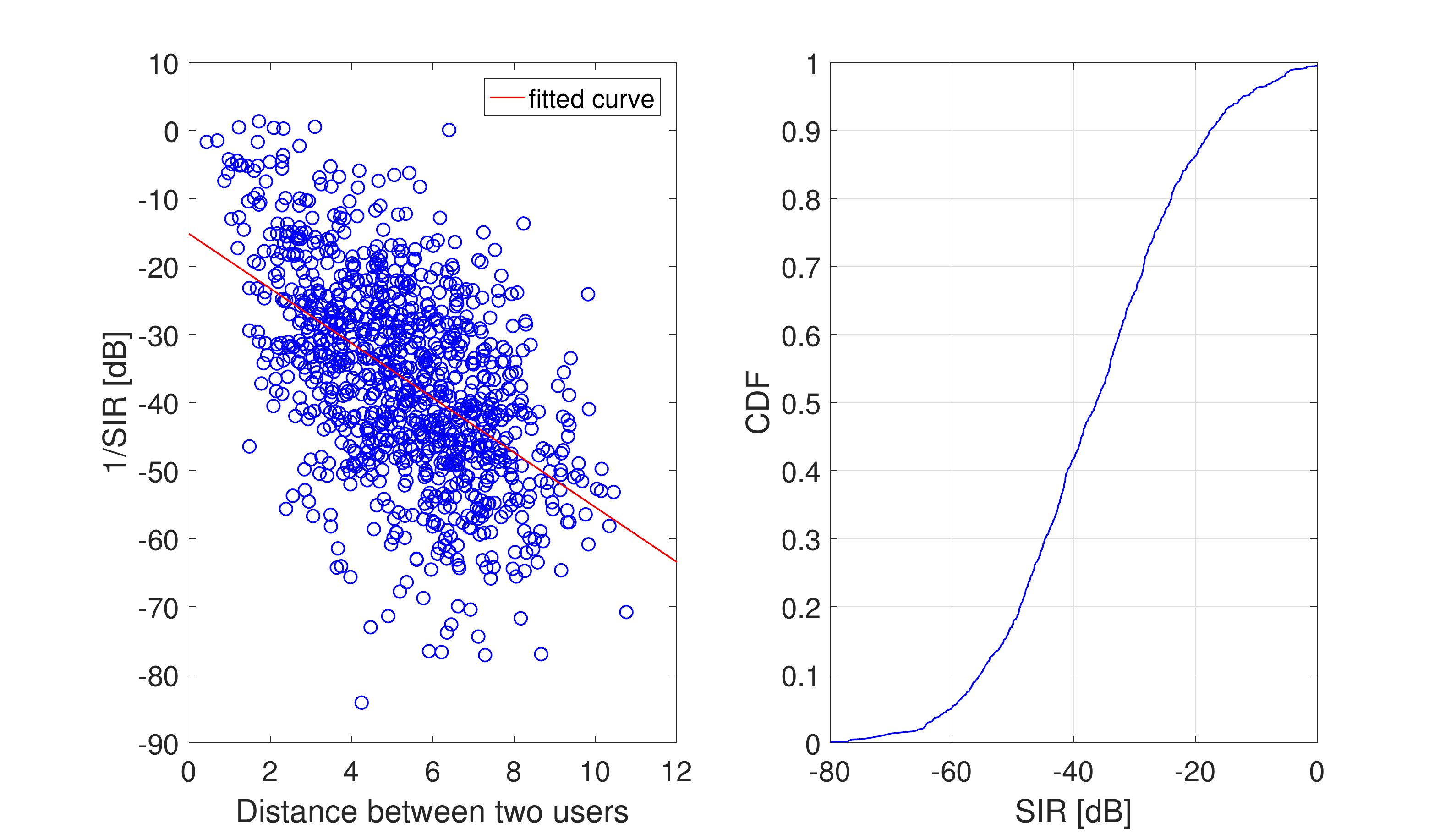}}
\vspace*{-7mm}
\caption{\label{fig3}The interference powers normalized by the signal powers, i.e., 1/SIR, are measured with two users in front of a square LIS-Unit with $L\!=\!0.5$ whose center is $x\!=\!y\!=\!z\!=\!0$. The locations of the two users are drawn from a uniform distribution inside a cube with $-4\!\leq\!x, y\leq\!4$ and $0\!<\!z\leq\!8$.}
\vspace*{-4mm}
\end{center}
\end{figure}

\begin{figure}
\begin{center}
\vspace*{-2.5mm}
\hspace*{-2mm}
\scalebox{0.285}{\includegraphics{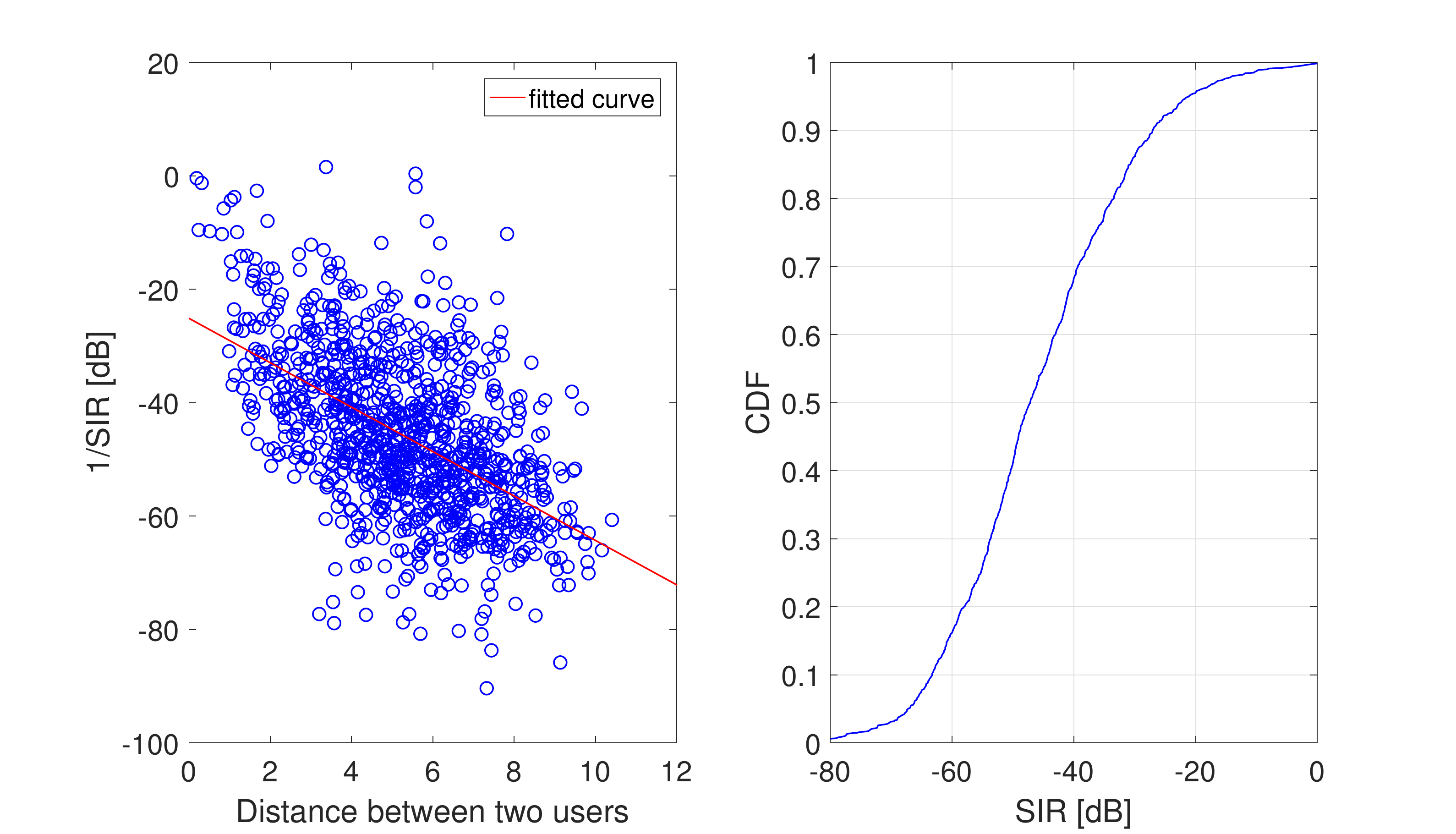}}
\vspace*{-7mm}
\caption{\label{fig4}Repeat of the tests in Fig. \ref{fig2} with $L\!=\!1$.}
\vspace*{-7mm}
\end{center}
\end{figure}

In Fig. \ref{fig3}, we evaluate the signal-to-interference ratio (SIR) for a case where $L\!=\!0.5$\footnote{Without explicitly pointed out, the unit of length, wavelength and coordinates are all in meter (m) in the rest of the paper.}, $\lambda\!=\!0.125$ (corresponding to a carrier-frequency 2.4 GHz) and two users that are uniformly located in front of the LIS-Unit\footnote{Assuming the LIS-Unit is implemented with discrete antenna-elements according to the sampling theory and the spacing between two adjacent antenna-elements is $\lambda/2$, then $L\!=\!0.5$ and $L\!=\!1$ corresponds to 64 and 256 antenna elements, respectively.}, with coordinates $-4\!\leq\!x, y\leq\!4$ and $0\!<\!z\leq\!8$ for both users. We test for 1000 realizations of random user locations and report the empirical cumulative probability density function (CDF). As can be seen, in almost 90\% of the test cases, the value of 1/SIR is below -20 dB, which shows that the interference from the other user is significantly suppressed. The same results can be seen from Fig. \ref{fig4} where we set a larger $L\!=\!1$, in which case the interference is further reduced and in almost 97\% of the test cases, the SIR is below -20 dB.

Utilizing the effectiveness of interference suppression with LIS-Unit, we next elaborate on optimal user assignments with the distributed LIS-system for sum-rate maximization and minimum user-rate maximization.

\section{RSS based User Assignments}

\subsection{Optimal user assignments with the LIS}

We consider a LIS deployment as in Fig. \ref{fig1}, where $M$ small LIS-units are deployed distributively forming a large LIS-system and serve $K$ ($K\!\leq\!M$) users simultaneously. The target is to find $K$ best LIS-Units that maximize the sum-rate and minimum user-rate, respectively, with each LIS-Unit serving a user separately.

Denote a set $\mathcal{P}$ comprising all the possible assignment schemes, whose cardinality equals  
\bea |\mathcal{P}|=\frac{M!}{(M-K)!}.\eea
Each assignment $\vec{p}\!\in\!\mathcal{P}$ contains $K$ elements with the $k$th element $p(k)$ denoting the index of the LIS-Unit which is selected to serve the $k$th user. The user-rate achieved with the $m$th LIS-Unit equals
\bea \label{Rk} R_k^{m}=\log\left(1+\frac{\big(\phi_{k,k }^{m}\big)^2}{N_0\phi_{k,k }^{m}
+\sum\limits_{\ell=0,\ell\neq k}^{K-1}|\phi_{k,\ell}^{m}|^2}\right)\!. \eea
Finding optimal assignments $p^{\ast}$ can be formulated as the following maximization problems, respectively:
{\setlength\arraycolsep{2pt} \bea \label{prob1} &&\text{Sum-rate:\; \;\quad\quad\qquad}p^{\ast}=\argmax_{\vec{p}\in\mathcal{P}} \sum_{k=0}^{K-1}R_k^{p(k)}, \\
\label{prob2} &&\text{Minimum user-rate:\; \;} p^{\ast}=\argmax_{\vec{p}\in\mathcal{P}} \left(\min_{k}\left(R_k^{p(k)}\right)\!\right)\!. \qquad\;  \eea}
\hspace{-1.2mm}The optimizations (\ref{prob1}) and (\ref{prob2}) can be solved in a brute-force manner for a small $M$. However, when $M$ and $K$ are large values, the complexity becomes prohibitive, which is not only because of the cardinality of $\mathcal{P}$, but also the computations for evaluating all $\phi_{k,\ell}^m$ that needs $MK^2$ operations in (\ref{phi}). Therefore, suboptimal user assignment algorithms are needed to simplify the complexities. First, we introduce the RSS based user assignment to reduce of complexity of evaluating $\phi_{k,\ell}^m$.

\subsection{RSS based user assignment}
According to the results shown in Sec. II-B, the interference power $|\phi_{k,\ell}^m|^2$ when $\ell\!\neq\!k$ is negligible compared to the signal power $|\phi_{k,k}|^2$ at each of the LIS-Unit. Therefore, the interference terms $|\phi_{k,\ell}^m|^2$ can be ignored in (\ref{Rk}). Then, maximizing $R_k^m$ is equivalent to maximizing $\phi_{k,k}^m$, and the optimization problems can be transferred to the following simplified problems:
{\setlength\arraycolsep{2pt} \bea \label{prob11} &&\text{Sum-rate: \;\;\quad\quad\qquad} p^{\ast}=\argmax_{\vec{p}\in\mathcal{P}} \sum_{k=0}^{K-1}\phi_{k,k}^{p(k)}, \\
\label{prob22} &&\text{Minimum user-rate: \;\;} p^{\ast}=\argmax_{\vec{p}\in\mathcal{P}} \left(\min_{k}\left(\phi_{k,k}^{p(k)}\right)\!\right)\!. \qquad\;  \eea}
\hspace{-1.4mm}As only $\phi_{k,k}^m$ are needed in (\ref{prob11}) and (\ref{prob22}), the complexity of evaluating $\phi_{k,k}^m$ reduces to $MK$ operations. 

Note that, although (\ref{prob11}) and (\ref{prob22}) are formulated from the LoS scenario, since $\phi_{k,k}^m$ denotes the RSS at the LIS-Unit for each user, the optimizations (\ref{prob11}) and (\ref{prob22}) also valid for scattering scenarios as explained in the next subsection.

\subsection{The validity of RSS based user assignments with reflections}
\begin{figure}[b]
\begin{center}
\vspace*{-2mm}
\hspace*{-5mm}
\scalebox{0.74}{\includegraphics{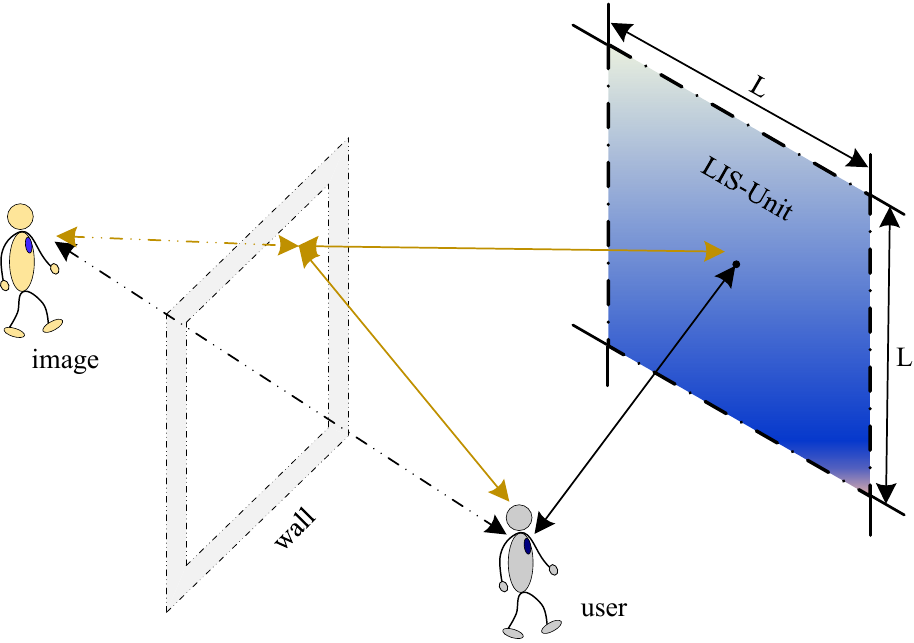}}
\vspace*{-2mm}
\caption{\label{fig5}An indoor scenario where a user communicating to a LIS-Unit with reflections from the wall received by the LIS-Unit.}
\vspace*{-1mm}
\end{center}
\end{figure}

In scattering environments, the RSS $\phi_{k,k}^m$ comprises the signals reflected from different scatters that reach the LIS-Unit. As we are considering a narrowband system, time-differences of signals from different scatters when arriving the LIS-Units are negligible. In Fig. \ref{fig5}, we depict an indoor scenario where besides the LoS signals from the user, there are also signals reflected by a surface that reach the LIS-Unit at the same time. In an ideal case, the reflected signals can be viewed as LoS signals from an image of the user created by the reflecting surface, with additional transmit power loss caused by attenuation\cite{IY02}. 

An important observation from Fig. \ref{fig5} is that, based on the interference suppressing with the LIS-Unit with MF procedure, the effective channel from the user and the image user are almost orthogonal. In more complex scattering environments with many scattering clusters, signal components from different clusters \cite{GO13} can still be assumed to be orthogonal which results in coherent RSS addition at the LIS-Units. Therefore, the RSS $\phi_{k,k}^m$ is still a valid measurement for representing the user-rate that can be achieved in scattering environments.

\subsection{Transferred linear assignment problems}
\begin{figure}[t]
\begin{center}
\vspace{1mm}
\hspace{4mm}
\scalebox{2}{\includegraphics{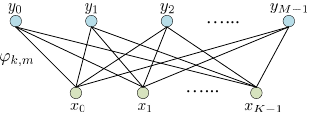}}
\vspace{-2mm}
\caption{\label{graph}A weighted bipartite graph $\vec{G}$ between the users and LIS-Units for solving the linear assignment problems (\ref{prbm1}) and (\ref{prbm2}), with the weight of each edge $(x_k, y_m)$ equals $\varphi_{k,m}$.}
\vspace{-7mm}
\end{center}
\end{figure}

With the complexity of evaluating $\phi_{k,\ell}^m$ reduced from $MK^2$ to $MK$, in a second step we reduce the complexity caused by the cardinality of $\mathcal{P}$ from $M!/(M-K)!$ to $\mathcal{O}(KM^2)$. We first introduce a coefficient matrix $\vec{W}$ such that the element $w_{k,m}\!=\!1$ if the $k$th user is assigned to the $m$th LIS-Unit; otherwise $ w_{k,m}\!=\!0$. In order to form LAPs, we define a cost matrix $\vec{\varPhi}$ with elements
\bea  \varphi_{k,m}= -\phi_{k,k}^{m}\leq 0.\eea
Then, the optimization problems (\ref{prob11}) and (\ref{prob22}) can be equivalently written as linear sum and bottleneck assignment problems, respectively, as:
{\setlength\arraycolsep{2pt} \bea \label{prbm1} &&\text{LSAP: \,\;\;} \min\quad \sum_{k=0}^{K-1}\sum_{m=0}^{M-1}w_{k,m}\varphi_{k,m}    \\
 \label{prbm2} &&\text{LBAP:\quad} \min\quad \max_{k}\left(\sum_{m=0}^{M-1}w_{k,m}\varphi_{k,m}\right)\! \\
&&\label{const1} \text{\,s.t.\;\;}   \sum_{k=0}^{K-1}w_{k,m}\leq1,\quad 0\leq m<M   \\
   &&\label{const2} \qquad\sum_{m=0}^{M-1}w_{k,m}=1, \quad 0\leq k<K   \\
  &&\label{const3} \qquad w_{k,m} \in\{0,1\}, \quad 0\leq k<K,\; 0\leq m<M .  \qquad\eea}
\hspace{-1.4mm}The optimizations in (\ref{prbm1}) and (\ref{prbm2}) under constraints (\ref{const1})-(\ref{const3}) can be efficiently solved through graph based algorithms. We first construct a weighted bipartite graph $\vec{G}\!=\!(\vec{V},\vec{E})$ as depicted in Fig. \ref{graph} in the following manner. We let $\vec{x}\!=\![x_0, x_1,\dots,x_{k-1}]$ with each vertex $x_k$ representing the $k$th user, and $\vec{y}\!=\![y_0, y_1,\dots,y_{M-1}]$ with each vertex $y_m$ representing the $m$th LIS-Unit. Hence, it holds that $\vec{x}\cap\vec{y}\!=\!\emptyset$, and we set $\vec{V}\!=\!\vec{x}\cup\vec{y}$ and $\vec{E}\!\subseteq\!\vec{x}\times\vec{y}$ is the set of possible matchings. The weight of each edge $(x_k, y_m)$ in $\vec{G}$ is $\varphi_{k,m}$. 

Following conventional notations of LAP, we make the following notations. Define labels $\ell(u)$ for each vertex in graph $\vec{G}$, and a feasible labeling $\ell\!:\!\vec{V}\!\to\!\mathbb{R}$ satisfies the condition $\ell(x_k)\!+\!\ell(y_m)\!\geq\!\varphi_{k,m}$. The neighborhood of a vertex $x_k$ is the set $N_\ell(x_k)\!=\!\{y_m:\ell(x_k)\!+\!\ell(y_m)\!=\!\varphi_{k,m}\}$ with all vertices $y_m$ that share an edge with $x_k$, and the neighborhood of a set $\mathcal{S}$ is $N_\ell(\mathcal{S})\!=\!\cup_{x_k\in\mathcal{S}}N_\ell(x_k)$. Let $\mathcal{P}$ be a matching of $\vec{G}$. A maximum matching is a matching $\mathcal{P}$ such that any other matching $\mathcal{P}\rq{}$ satisfies $|\mathcal{P}\rq{}|\!\leq\!|\mathcal{P}|$. A perfect matching is a matching $\mathcal{P}$ in which every vertex in $\vec{x}$ is adjacent to some edge in $\mathcal{P}$. A vertex $y_m$ is matched if it is endpoint of edge in $\mathcal{P}$, otherwise it is unmatched. 

With the above notations, we can apply the Kuhn-Munkres algorithm \cite{KMA, BL71, G06} to find a perfect matching $\mathcal{P}$, i.e., solutions of $w_{k,m}$ for LSAP (\ref{prbm1}), which is guaranteed to reach a global optimal with a numerical complexity $\mathcal{O}(KM^2)$ and summarized in Algorithm 1.

\begin{algorithm}[ht!]
\vspace{2mm}
	\caption{Kuhn-Munkres algorithm for solving (\ref{prbm1}).}
	\label{alg:1}
      \begin{algorithmic}[1]
      \STATE Initialize $\ell(y_m)\!=\!0, \forall y_m\!\in\!Y$, $\ell(x_k)\!=\!\max\limits_{m}(\varphi_{k,m})$, and $\mathcal{P}\!=\!\emptyset$. \\
       \STATE If $\mathcal{P}$ is perfect, stop and return $\mathcal{P}$; otherwise select an unmatched vertex $x_k\!\in\!X$ and set $\mathcal{S}\!=\!x_k$, $\mathcal{T}\!=\!\emptyset$.  \\
       \STATE If $N_\ell(\mathcal{S})\!=\!\mathcal{T}$, update labeling according to:
       {\setlength\arraycolsep{2pt} \bea \Delta=\min_{x_k\in\mathcal{S},\; y_m\notin\mathcal{T}}\left(\ell(x_k)+\ell(y_m)-\varphi_{k,m} \right),\notag \\
       \ell(u)=\left\{\begin{array}{cc}\ell(u)-\Delta, &u\in\mathcal{S} \\  \ell(u)+\Delta, &u\in\mathcal{T} \\ \ell(u), &\mathrm{otherwise}. \end{array}\right. \eea}
       
       \STATE  If $N_\ell(\mathcal{S})\!\neq\!\mathcal{T}$, select $y_m\!\in\! N_\ell(\mathcal{S})\!-\!\mathcal{T}$: If $y_m$ free, $x_k-y_m$ is an augmenting path. Augment $\mathcal{P}$ and go to Step 2. If $y_m$ is matched to some $x_j$, extend alternating tree: $\mathcal{S}\!=\!\mathcal{S}\!\cup\!x_j$ and $\mathcal{T}\!=\!\mathcal{T}\!\cup\!y_m$, and then go to Step 3.\\
\end{algorithmic}
\end{algorithm}

Further, for LBAP (\ref{prbm2}) we can also apply the Threshold Algorithm \cite[Ch. 6]{BL71} to find an optimal solution which is summarized in Algorithm 2, and has a similar complexity \cite{DM99} as Algorithm 1. When $M$ and $K$ are large values, the save of computational costs of Algorithm 1 and 2 compared to the brute-force method is significant.

\section{Numerical results}
In this section, simulation results are provided for evaluating the optimal user assignments with LSAP (\ref{prbm1}) and LBAP (\ref{prbm2}). The optimal user assignments obtained with brute-force methods to solve the original problem (\ref{prob1}) and (\ref{prob2}) are also presented as upper bounds. Further, we average the rates obtained with all the possible user assignments with size $|\mathcal{P}|$ to serves as lower bounds, which reflect the rates achieved with random user assignments. We consider both LoS and scattering scenario, and put a special interested in the rates can be achieved per m$^2$ deployed surface-area of each LIS-Unit and per user. In all the tests, we assume all users transmitting at a transmit power $P\!=\!20$ dB and a noise power density $N_0\!=\!1$, if not otherwise explicitly pointed out. 

We assume that all LIS-Units have identical square shapes with identical surface-area $L\!\times\!L$ m$^2$ each, and are deployed close to each other to form a large LIS-system with centers uniformly distributed along the line $y\!=\!z\!=\!0$, and with the middle LIS-Unit centered at location (0,0,0). Further, we assume that the users are uniformly distributed in front of the LIS-Units with coordinates $(x, y, z)$ satisfying $-2\!\leq\!x\!\leq\!2$, $-2\!\leq\!y\!\leq\!2$, and $0\!<\!z\!\leq\!4$. For all the test scenarios, we generate 2000 realizations of random user locations.

Note that, although Algorithm 1 and 2 are guaranteed to reach optimal solutions for solving (\ref{prbm1}) and (\ref{prbm2}), there are still rate-losses compared to the brute-force algorithms due to the simplification of using RSS in (\ref{prob11})-(\ref{prob22}) to replace the capacities in (\ref{prob1})-(\ref{prob2}).

\begin{algorithm}[b]
\vspace{2mm}
	\caption{Threshold algorithm for solving (\ref{prbm2}).}
	\label{alg:2}
      \begin{algorithmic}[1]
      \STATE Initialize $\varphi^{\ast}\!=\!\max\limits_m(\min\limits_k\varphi_{k,m},\min\limits_m\varphi_{k,m})$ and $\mathcal{P}\!=\!\emptyset$. \\
       \STATE Define a bipartite graph $\vec{G}(\varphi^{\ast})$ whose edges correspond to $\varphi_{k,m}\!\leq\!\varphi^{\ast}$.  \\
       \STATE Find a maximum matching $\mathcal{P}$ in $\vec{G}(\varphi^{\ast})$. If $|\mathcal{P}|\!=\!K$, stop and return $\mathcal{P}$; otherwise go to Step 4. \\
       \STATE Find a minimal row and column covering of the elements $\varphi_{k,m}\!\leq\!\varphi^{\ast}$ in $\vec{\varphi}$. Set $\varphi^{\ast}$ to the minimum of the remaining elements in $\vec{\varphi}$ after removing the rows and columns from the minimal covering, and then go to Step 2.\\
\end{algorithmic}
\end{algorithm}

\subsection{The user assignments in LoS scenario}
\begin{figure}[t]
\begin{center}
\vspace*{-0.5mm}
\hspace*{-2mm}
\scalebox{0.31}{\includegraphics{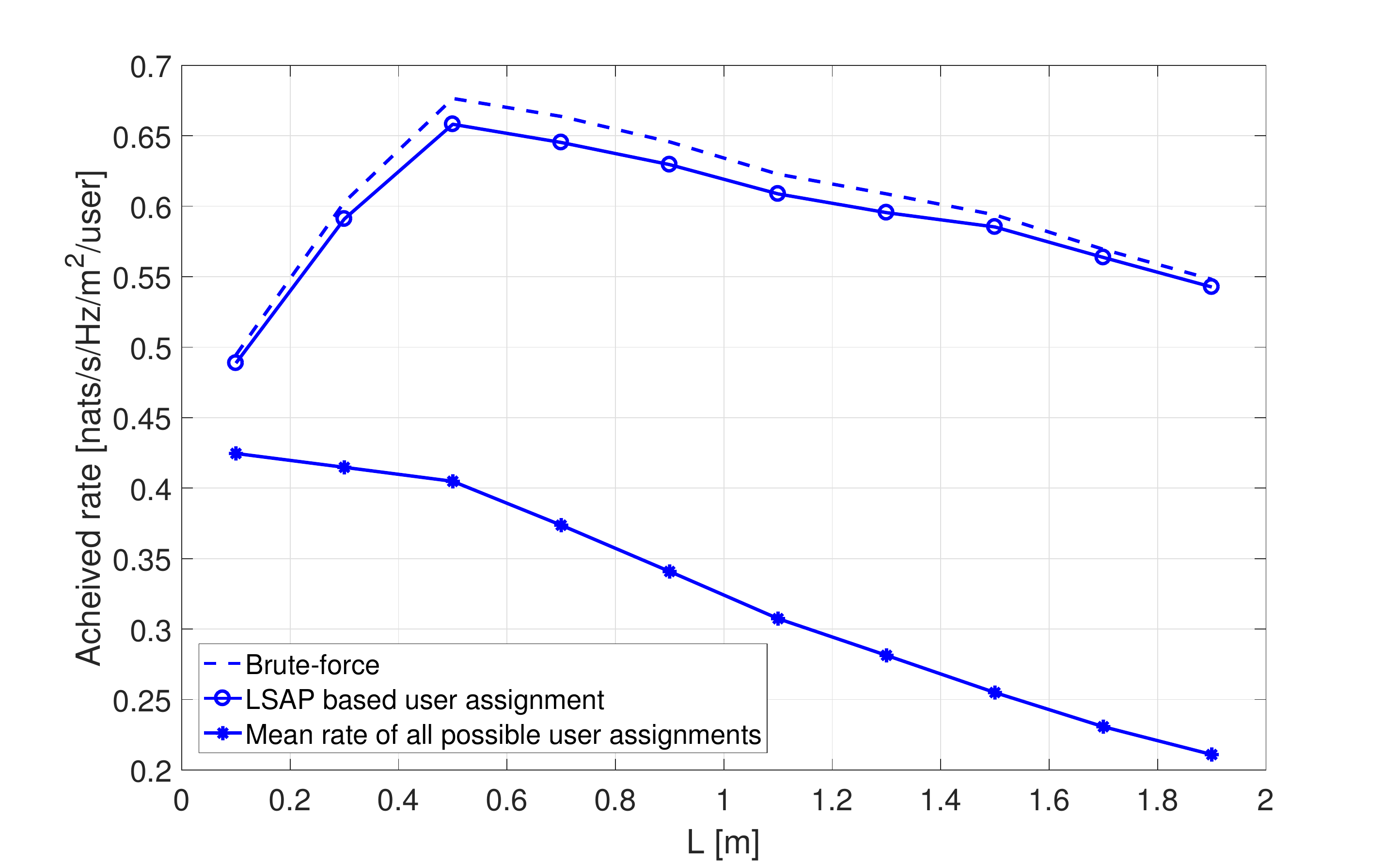}}
\vspace*{-8mm}
\caption{\label{fig7}Sum-rate maximization in LoS scenario with 6 LIS-Units and 2 users.}
\vspace*{-4mm}
\end{center}
\end{figure}

First, we evaluate the user assignment in LoS scenarios, and consider a LIS-system with $M\!=\!7$ LIS-Units deployed on a plane. The sum-rate maximized with solving LSAP (\ref{prbm1}) and the brute-force search over (\ref{prob1}) are presented in Fig. \ref{fig7}, after normalizing by the surface-area of a LIS-Unit and the number of users. As can be seen, the proposed user assignments is quite close to the optimal, and much better than random user assignments. Furthermore, the achieved rate per user first increases when the surface-area increases and then starts to decrease. This is essentially because that, as the surface-area initially increases, the inter-user interference suppression improves which in turn improves the achieved rate of each user. After the inter-user interference suppression becomes perfect, further increasing the surface-area of each LIS-Unit decreases the achieved rate per m$^2$ deployed surface-area as the edge parts of the LIS-Unit provide lower rates compared to the central parts of the LIS-Unit.

\begin{figure}
\begin{center}
\vspace*{-3mm}
\hspace*{-2mm}
\scalebox{0.31}{\includegraphics{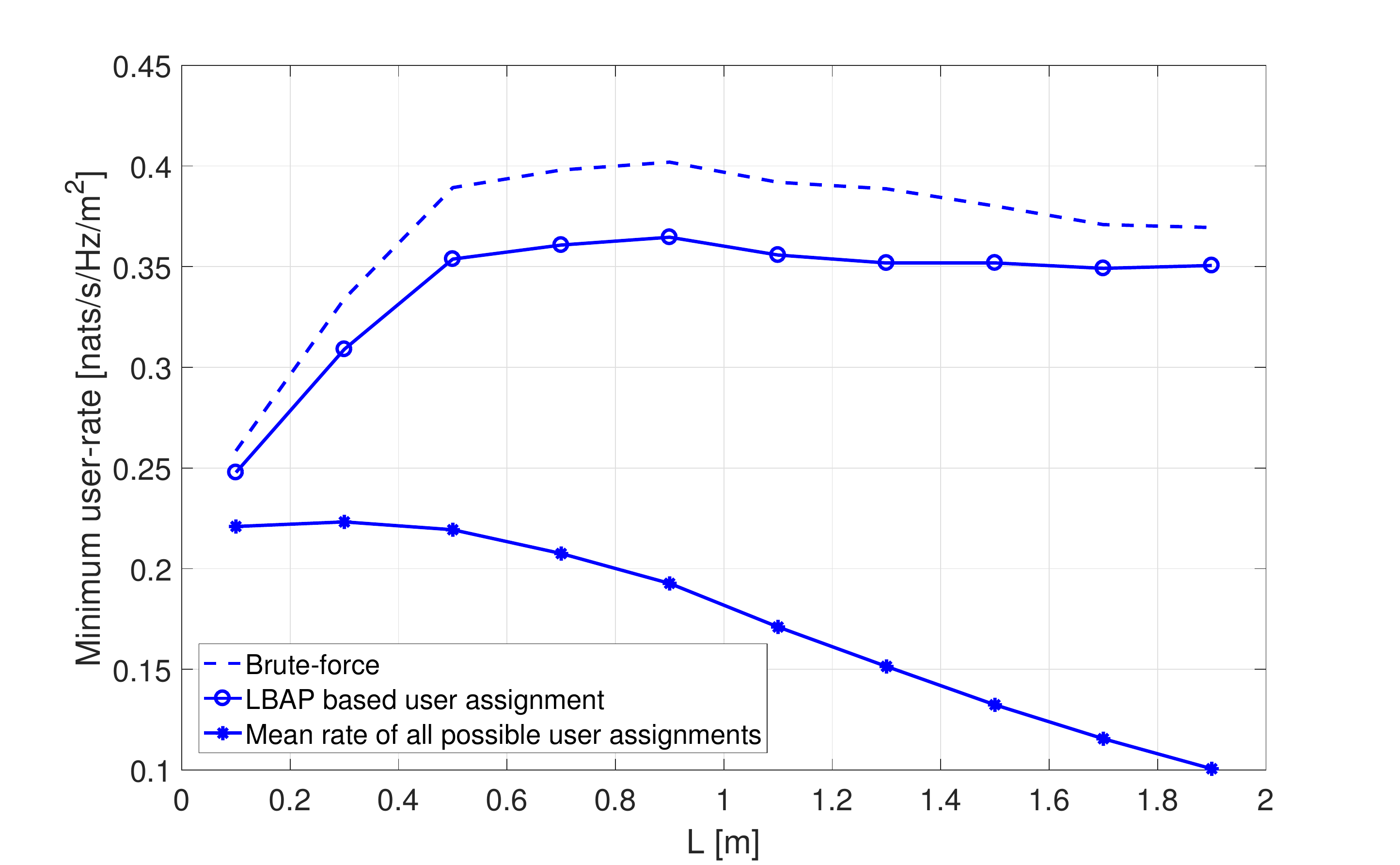}}
\vspace*{-8mm}
\caption{\label{fig8}Repeat the tests in Fig. \ref{fig7} for minimum user-rate maximization.}
\vspace*{-7mm}
\end{center}
\end{figure}

In Fig. \ref{fig8}, the minimum user-rate maximized with solving LBAP (\ref{prbm2}) and the brute-force search over (\ref{prob2}) are presented. As can be seen, the proposed user assignments has $\sim$10\% minimum user-rate losses compared to the brute-force search in this case, due to the fact that, the minimum user-rate is more sensitive to the user arrangement than the sum-rate. However, the proposed user assignment has a much lower complexity than the brute-force, and it still significantly outperforms the random user assignments. 

\subsection{The user assignments with reflections}

Next, to evaluate the user assignment performance in scattering environments, we consider a LIS-system with $M\!=\!5$ LIS-Units deployed on the front wall in a hall, and with $N\!=\!2$ users randomly located inside the hall. In addition, we consider the wall reflections and create 5 images for each user corresponding to the 5 walls except the front wall where the LIS-Units are deployed according to Fig. \ref{fig5}. The attenuation is assumed to be $-3$ dB for all walls.

\begin{figure}[t]
\begin{center}
\vspace*{-0.5mm}
\hspace*{-2mm}
\scalebox{0.31}{\includegraphics{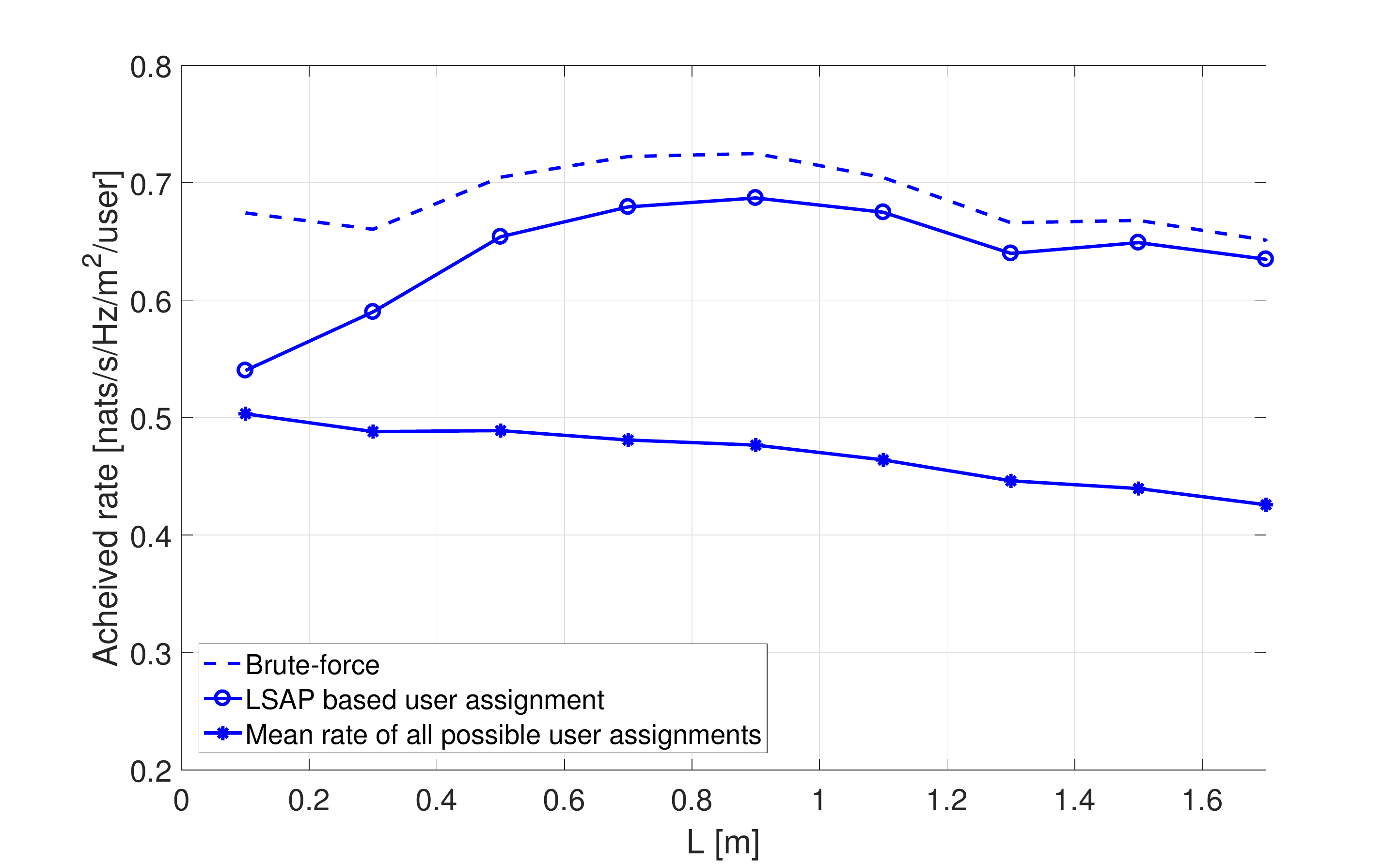}}
\vspace*{-8mm}
\caption{\label{fig9}Sum-rate maximization under wall-reflections scenario with 5 LIS-Units and 2 users.}
\vspace*{-4mm}
\end{center}
\end{figure}

In Fig. \ref{fig9}, the normalized sum-rates are presented, and as can be seen, the conclusions are similar to those drawn from LoS scenarios except for the cases that $L$ is really small. This is essentially because that, with small $L$ such as  $L\!\leq\!0.3$, the inter-user interference suppression is not as good as LoS scenarios due to the presented 5 images for each user. Nevertheless, for relatively large $L$ it can be seen that, the proposed user assignments still work well compared to the brute-force results.

\begin{figure}
\begin{center}
\vspace*{-3mm}
\hspace*{-2mm}
\scalebox{0.31}{\includegraphics{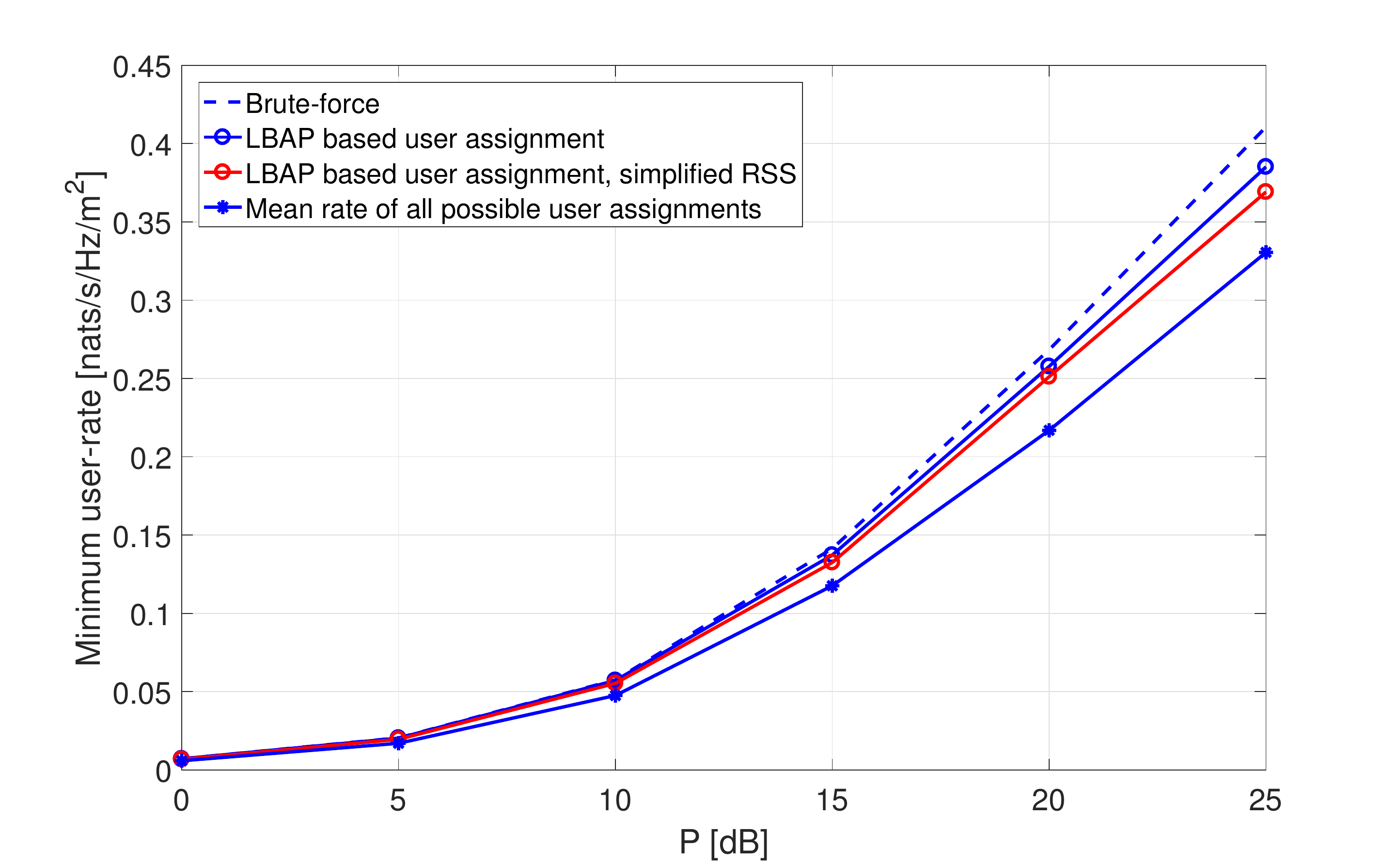}}
\vspace*{-8mm}
\caption{\label{fig10}Minimum user-rate maximization with measuring RSS only at the center of each LIS-Unit in a LoS scenario with 5 LIS-Units and 2 users.}
\vspace*{-7mm}
\end{center}
\end{figure}

\subsection{An extension of the RSS based user arrangements}
At last we discussion an extension of the proposed RSS based algorithms. Instead of measuring the RSS $\phi_{k,k}^m$ over the whole LIS-Unit for each user, a reduced complexity RSS measurement is to only measure the RSS at finitely many discrete positions. An ultimate simplified RSS measurement algorithm would be only evaluate the RSS at the center of each LIS-Unit, which we denote as $\tilde{\phi}_{k,k}^m$ and is calculated according to (\ref{phi1}) as
\bea \label{tphi} \tilde{\phi}_{k,k}^m=\frac{z_k}{4\pi\eta_k^{\frac{3}{2}}}\propto z_k \eta_{k,m}^{-\frac{3}{2}}, \eea
where $\eta_{k,m}$ is the square of distance between the $k$th user and the $m$th LIS-Unit. Clearly, replacing $\phi_{k,k}^m$ by $\tilde{\phi}_{k,k}^m$ would degrade the performance of user assignment. However, under the cases that the users are far away from the LIS-Unit or the surface-area of the LIS-Unit is sufficiently small compared to the distances from the users to the LIS-Unit, $\tilde{\phi}_{k,k}^m$ (multiplying with the surface-area of the LIS-Unit) is a good estimate of $\phi_{k,k}^m$. Therefore, it is of interest to evaluate the performance degradation in general cases.

In Fig. \ref{fig10}, we evaluate the minimum user-rate maximizations with the reduced-complexity RSS measurement. We test a similar case as in Fig. \ref{fig8} with $M\!=\!5$ LIS-Units with $L\!=\!0.2$ and $N\!=\!2$ users. As can be seen, with simplified RSS $\tilde{\phi}_{k,k}^m$, the minimum user-rate is slightly lower then with the original RSS $\phi_{k,k}^m$ for large transmit power. The sum-rate maximization is also evaluated for the simplified RSS, but as the achieved rates are almost the same with $\tilde{\phi}_{k,k}^m$ and $\phi_{k,k}^m$, they are not presented. Nevertheless, the results in Fig. \ref{fig10} shows the potential to evaluate the RSS at a number of sampled discrete points on the LIS-Unit (or even only at the center), which can significantly simplify the complexity of computing $\phi_{k,k}^m$.

\section{Summary}

We have considered optimal user assignments for a distributed large intelligent surface (LIS) system with $M$ separate LIS-Units. The objective is to select $K$ best LIS-Units to serve $K$ ($K\!\leq\!M$) autonomous users simultaneously. By constructing a cost matrix based on the received signal strength (RSS) at each LIS-Unit for each user, we obtain a weighted bipartite graph between the $K$ users and $M$ LIS-Units. Utilizing the effectiveness of LIS-Unit in inter-user interference suppression and with the constructed bipartite graph, the optimal user assignments for sum-rate maximization can be transferred to a linear sum assignment problem (LSAP), and for the minimum user-rate maximization can be transferred to a linear bottleneck assignment problem (LBAP), respectively. The linear assignment problems (LAPs) are then solved through the classical Kuhn-Munkres and Threshold algorithms with time complexities $\mathcal{O}(KM^2)$. We show through numerical results that, the proposed user assignments perform close to the optimal assignments both for considered line-of-sight (LoS) and scattering environments.

\end{document}